\documentclass[letterpaper]{article}
\usepackage{aaai}
\usepackage{times}
\usepackage{enumitem}
\usepackage{hyperref}
\usepackage{helvet}
\usepackage{courier}
\usepackage{graphicx} 
\usepackage{array}
\usepackage{wrapfig}
\usepackage{multirow}
\usepackage{subcaption}
\usepackage{xspace}
\usepackage{tabu}
\graphicspath{ {/} }
\newcommand{\reprowd}{\textsc{Reprowd}\xspace}

\begin{document}
%
\title{Reprowd: Crowdsourced Data \\Processing Made Reproducible}
\author{Ruochen Jiang, Jianan Wang\\
Simon Fraser University\\
Burnaby, Vancouver, Canada\\
\{ruochenj, jnwang\}@sfu.ca\\
}
\maketitle

\section{Introduction}

\noindent 

Crowdsourcing is a multidisciplinary research area including disciplines like artificial intelligence, human-computer interaction, database, and social science. One of the main objectives of AAAI HCOMP conferences is to bring together researchers from different fields and provide them opportunities to exchange ideas and share new research results. To facilitate cooperation across disciplines, \emph{reproducibility} is a crucial factor, but unfortunately it has not gotten enough attention in the HCOMP community.

Imagine a researcher Bob did a crowdsourcing experiment, and another researcher Ally would like to reproduce the experiment (Note that in this paper, we trust the crowdsourced answers collected by Bob, which is a weaker claim of reproducibility than~\cite{DBLP:conf/www/Paritosh12}.). This reproducibility process could take a lot of time for both Bob and Ally. From the Bob's perspective, he has to spend additional time in modifying the code since the code written for doing the experiment is different from that used for reproducing the experiment. For example, the former requires to collect answers from crowd workers, but the latter just reuses the cached crowdsourced answers. From the Ally's perspective, once she receives the modified code and the crowdsourced answers, she might find it hard to examine the experimental result since the code may not be easy to extend or the crowdsourced answers may not contain enough lineage information (e.g., when were the tasks published? which workers did the tasks?).

These issues will discourage researchers from sharing experimental results or analyzing others' results to derive new insights, resulting in a significant negative impact on the HCOMP community. While some recently developed tools~\cite{DBLP:conf/uist/LittleCGM10,DBLP:conf/sigmod/ChirigatiRSF16,DBLP:conf/hcomp/SheshadriL13} can be used to mitigate the impact of these issues, they do not fully meet Bob and Ally's requirements. Reprozip~\cite{DBLP:conf/sigmod/ChirigatiRSF16} is tool for automatically packing an experiment along with the entire programming environment (e.g., dependent libraries or packages). It saves the time for deploying an experiment but not the time that Bob spends in modifying the code or Alley spends in examining the code. Turkit~\cite{DBLP:conf/uist/LittleCGM10} proposes a crash-and-rerun programming model. This programming model can help Bob to solve his issue, but add extra burden to Ally. Since Turkit caches the function's returned values into a database \emph{in sequence}, Ally has to be very careful with the order of these function calls. If she accidentally swapped the order of two functions or added a new function between them, the whole experiment would break. This restriction make it even harder for Ally to examine the Bob's experimental result. 

In this paper, we present \reprowd, a system aiming to address these issues. We identify two requirements for making a crowdsourcing experiment easy to reproduce. 1) Sharable. Once Bob finishes a crowdsourcing experiment, he should be able to directly share the experiment to Ally without any need to change the code. (2). Examinable. The experiment should capture complete lineage information about crowdsourced answers and allow Ally to extend the code more easily. 

Due to our database background, we restrict the \reprowd's current focus on the database field only. Most of the crowdsourcing works in the database field are centered around the implementations of crowdsourced data processing operators~\cite{li2016crowdsourced}. That is, how to combine computers and crowds to implement traditional database operators such as join, sort, and max. Despite the restricted focus, \reprowd could be beneficial to any research field that needs to collect data from the crowd.

A key insight in designing \reprowd is to model a list of steps for doing a crowdsourcing experiment as a sequence of manipulations of a tabular dataset called \emph{CrowdData}. This idea enables us to leverage some existing techniques that were originally developed for data management, such as data recovery and data lineage, to address reproducibility challenges. Specifically, in order to satisfy the ``sharable" requirement, the system guarantees that any manipulation of CrowdData is fault recovery. That is, when the program is crashed, rerunning the program is as if it has never crashed. Thus, at any given point, Ally can simply rerun the Bob's code to reproduce his experimental result. In order to satisfy the ``examinable" requirement, CrowdData not only contains complete lineage information about crowdsourced answers but also allows other researchers to extend the code using the provided APIs. We find that the CrowdData abstraction is general enough to be used in re-implementing a large number of crowdsourced data processing algorithms in the literature. We have implemented two crowdsourced join algorithms~\cite{DBLP:journals/pvldb/WangKFF12,DBLP:conf/sigmod/WangLKFF13} and shown that these algorithms can inherit the sharable and examinable requirements from CrowdData for free. In the future, we will continue to add more algorithms into the system.

\section{System Overview}

\begin{figure}[t]
    \centering
    \includegraphics[scale = 0.6]{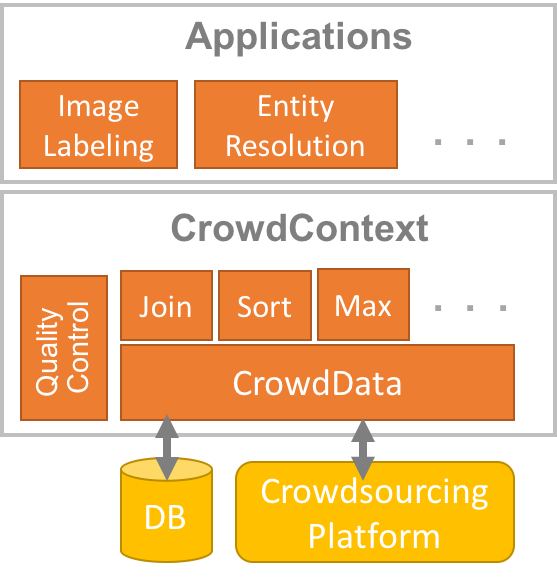}
    \caption{\reprowd Architecture}
    \label{fig:arch}\vspace{-1em}
\end{figure}

Figure~\ref{fig:arch} shows the  \reprowd architecture. Central to the system is CrowdData, which serves as a bridge to connect high-level crowdsourced operators (e.g., join) with an underlying crowdsourcing platform and database. The quality control component implements a number of widely used techniques for improving the quality of crowdsourced answers. All of these components are encapsulated into CrowdContext, the main entry point for \reprowd functionality. We also implement a few example applications such as image labeling and entity resolution for new users to try out the system.

\begin{figure}[t]
    \centering
    \includegraphics[scale = 0.55]{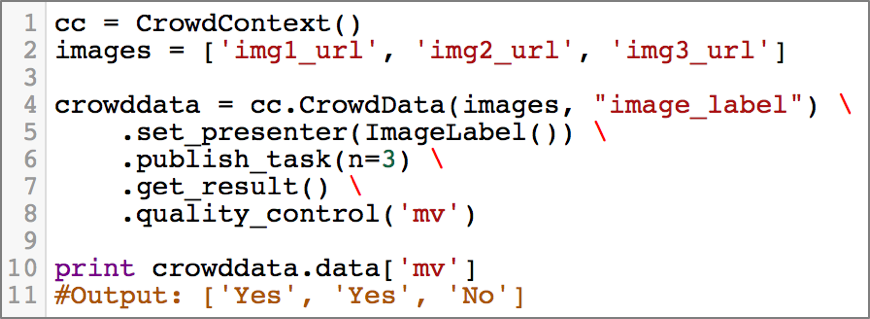}
    \caption{Bob's code for doing a crowdsourcing experiment (Label three images and use MV for quality control.). Ally can rerun the code to reproducing the experimental result.}
    \label{fig:bob-code}\vspace{1em}
\end{figure}

Next we use a simple example to explain how \reprowd works. Suppose Bob wants to label three images, where the label of an image is either ``Yes" or ``No". Each image will be assigned to three workers and then uses Majority Vote (MV) to decide its final label.
Figure~\ref{fig:bob-code} shows the Bob's code for doing this experiment. It consists of five steps: 

\begin{enumerate}[topsep=0pt, leftmargin=2\parindent,itemsep=-0.25em]
    \item Preparing input data (Line~4)
    \item Choosing a web user interface (Line~5)
    \item Publishing tasks to a crowdsourcing platform (Line~6)
    \item Getting results from the platform (Line~7)
    \item Using MV to determine the final labels (Line~8)
\end{enumerate}

\noindent Each step will be mapped to a manipulation of a tabular dataset. In step~1, the system initializes a table of two columns, where one is an \emph{id} column and the other is an \emph{object} column, and then populates the table with three rows: (1, ``img1\_url"), (2, ``img2\_url"), (3, ``img3\_url"). In step~2, it keeps the table unchanged. In step~3, it adds a new column \emph{task} to the table, which stores the related information about the published tasks. In step 4, it adds another new column \emph{result}, storing the related information about the crowdsourced answers. In step 5, it adds a column \emph{mv}, storing the majority-vote results. In addition, to make the data in the table fault recovery, the \emph{task} and \emph{result} columns will be stored persistently in a database, but the other columns will not be stored since they can be easily recovered through re-computation.

\begin{figure}[t]
    \hspace*{-1em}
    \includegraphics[scale = 0.46]{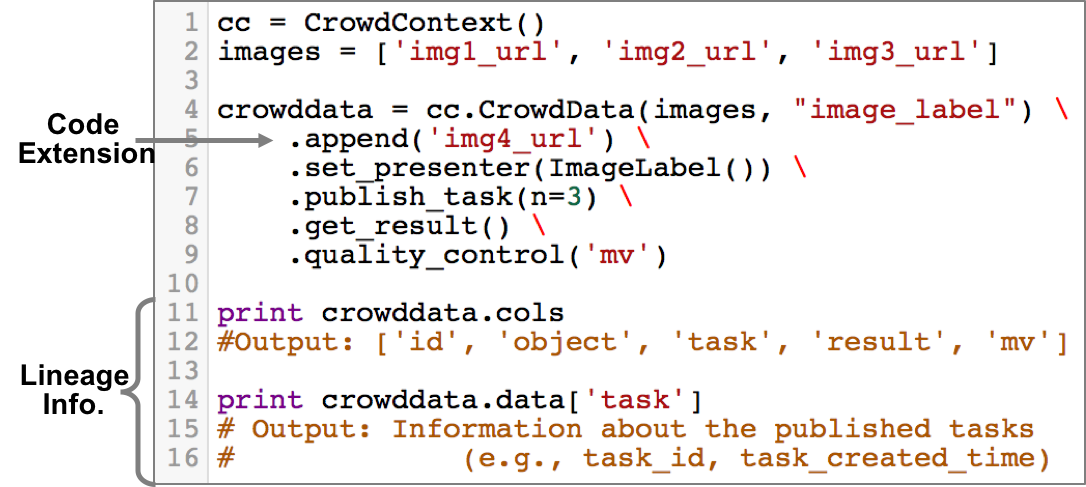}
    \caption{Ally's code for examining the experiment. She can extend the code by labeling more images or check the lineage information of the Bob's experiment.}
    \label{fig:ally-code}
\end{figure}

After Bob finishes the experiment, he can share the code along with the database file to Ally. Once Ally receives them, she can simply rerun the Bob's code to reproduce his experiment. Moreover, as shown in Figure~\ref{fig:ally-code}, she can further examine the experiment by building up a new experiment based on the Bob's (Line~5) or checking the lineage information of the experiment (Lines~11-16).

\section{Conclusion}

This paper points out that reproducibility is a key factor to facilitate research cooperation, and should get more attention in the HCOMP community. The paper presents \reprowd, a system aiming to make it easy to reproduce crowdsourced data processing research. A key innovative aspect of the system is the CrowdData abstraction, which maps a crowdsourcing experiment to manipulations of a tabular dataset. We describe how the mappings work, and explain why this new abstraction makes crowdsourcing experiments sharable and examinable. We have open sourced \reprowd at \url{http://sfu-db.github.io/reprowd/}. In the future, we will add the implementations of more crowdsourced data processing operators into the system.


\bibliography{reprowd.bib}
\bibliographystyle{aaai}

\end{document}